# Characterization of Age-related Microstructural Changes in Locus Coeruleus and Substantia Nigra Pars Compacta

Jason Langley[1], Sana Hussain[2], Justino J. Flores[3], Ilana J. Bennett[3], and Xiaoping Hu[1,2]*

1. Center for Advanced Neuroimaging, University of California Riverside, Riverside, CA
2. Department of Bioengineering, University of California Riverside, Riverside, CA
3. Department of Psychology, University of California Riverside, Riverside, CA

*Correspondence to: Xiaoping P. Hu (xph 'at' engr.ucr.ucr.edu)

**Abstract:**

Locus coeruleus (LC) and substantia nigra pars compacta (SNpc) degrade with normal aging, but not much is known regarding how these changes manifest in MRI images, or whether these markers predict aspects of cognition. Here, we use high-resolution diffusion-weighted MRI to investigate microstructural and compositional changes in LC and SNpc in young and aged cohorts, as well as their relationship with cognition. In LC, the older cohort exhibited a significant reduction in mean and radial diffusivity, but a significant increase in fractional anisotropy compared to the young cohort. We observed a significant correlation between the decrease in LC mean and radial diffusivities and measured examining delayed recall. This observation suggests that LC is involved in retaining cognitive abilities. In addition, we observed that iron deposition in SNpc occurs early in life and continues during normal aging. Since neuronal loss occurs in both LC and SNpc in Parkinson's disease, but occurs only in the LC in Alzheimer's disease, our results may lead to early stage imaging biomarkers for these diseases.

## 1. Introduction

Locus coeruleus (LC) and substantia nigra pars compacta (SNpc) are catecholamine nuclei situated in brainstem and subcortex, respectively. They both consist of neuromelanin generating catecholaminergic neurons and can be delineated in neuromelanin-sensitive MRI images [1-3]. Both nuclei are thought to play prominent roles in cognition. Specifically, LC plays a major role in arousal, attention modulation, and memory [4,5], and SNpc participates in motor function, novelty processing, and temporal processing [6]. Loss of melanized neurons in one or both nuclei is known to occur in age-related disorders, including Alzheimer's disease [7] and Parkinson's disease (PD) [8]. However, less is known about how these regions are affected in normal aging.

Histological studies have revealed that both LC and SNpc undergo neuronal loss or compositional changes in normal aging. Approximately 40% of LC neurons are lost by age 40 [9] while neuronal loss in SNpc is estimated to occur at a rate of approximately 4.7% per decade of life [10]. These microstructural differences can be examined *in vivo* with diffusion imaging, which measures the movement of molecular water [11]. Measures such as the degree of restricted diffusion (fractional anisotropy, FA) are sensitive to altered microstructural "integrity" (e.g., degeneration or demyelination), particularly in highly aligned white matter. The average (mean diffusivity, MD) or perpendicular (radial diffusivity, RD) rates of diffusion may instead capture microstructural differences in gray matter, such as the catecholamine nuclei of interest here. Previous studies in aging have shown changes in these diffusion metrics in deep gray matter structures (e.g., putamen) in older relative to younger [12].

To date, however, few studies have examined microstructural changes in substantia nigra localized by neuromelanin-sensitive MRI and no studies have assessed microstructural changes in LC using diffusion tensor imaging (DTI). One reason may be that assessment of catecholamine nuclei microstructure is hampered by the low resolution of diffusion-weighted images, which is typically insufficient to resolve LC. Furthermore, as SNpc experiences age-related iron deposition in addition to neuronal loss [13], SNpc diffusion metrics might be influenced by age-related iron accumulations in SNpc. Specifically, iron reduces diffusivity values [14] in diffusion-weighted acquisitions using monopolar diffusion encoding gradients [15].

Iron deposition can be measured using quantitative susceptibility mapping (QSM) or $R_2^*$ [16]. In substantia nigra, age-related changes in iron [17-20] and microstructure [12, 21] have been assessed using regions of interest (ROIs) placed in $T_2$-weighted images [17-20]. However, these ROIs are mostly spatially incongruent to the neuromelanin-sensitive substantia nigra (i.e. SNpc) [22] and, to date, age-related iron deposition or microstructural alterations have not been examined in SNpc defined by neuromelanin-sensitive MRI.

The current study takes advantage of recent advances in MRI hardware and acquisition strategies that have allowed for diffusion imaging with sub-millimeter in plane resolution, suitable for imaging LC. In this work, we will examine age-related microstructural and compositional differences in the catecholamine nuclei (LC, SNpc) using a combination of diffusion imaging and QSM/$R_2^*$. In addition, we assess the influence of iron on diffusion measures from diffusion-weighted sequences using monopolar and bipolar diffusion encoding gradients. The functional impact of these age-related structural differences are further ascertained in relation to memory performance.

## 2. Materials and Methods

Sixty-one subjects (24 aged and 37 young subjects) participated in this study. Subjects in the aged cohort were recruited from the Riverside community and subjects in the young cohort were recruited from the student population at University of California-Riverside. Participants were excluded from the study if there were any contraindications to MRI imaging or if they had a diagnosed neurological condition. All subjects participating in the study gave written informed consent in accordance with local institutional review board (IRB) regulations. One young subject was excluded from the diffusion-weighted MRI analysis due to problems with data acquisition and three subjects (2 aged and 1 young) were excluded from the analyses because of significant motion artifacts. The final sample size used in the analysis was 22 aged subjects and 35 subjects in the young cohort. Demographic information (gender, age) and Montreal cognitive assessment scoring (MOCA) and Rey Auditory Verbal Learning Test (RAVLT) scores were collected on each subject. Group means for Age, MOCA, and RAVLT scores are given in Table 1.

**Table 1.** Demographic information for the subject populations used in statistical analysis in this study. MOCA - Montreal cognitive assessment scoring; RAVLT - Rey Auditory Verbal Learning Test.

| Variable | Young (n=35) | Aged (n=22) | $p$ Value |
|---|---|---|---|
| Gender (M/F) | 15/20 | 9/13 | 0.414 |
| Age (yrs) | 20.7±2.2 | 73.0±6.7 | $<10^{-4}$ |
| MOCA score | 26.9±1.9 | 26.7±1.9 | 0.78 |
| RAVLT Total | 49.8±7.5 | 42.7±12.4 | 0.01 |
| RAVLT Immediate | 11.5±2.2 | 7.7±2.2 | $<10^{-4}$ |
| RAVLT Delay | 11.1±2.5 | 8.0±3.8 | 0.0007 |

*2.1 Image Acquisition*

Imaging data were acquired on a 3 T MRI scanner (Prisma, Siemens Healthineers, Malvern, PA) using a 32-channel receive only coil at the Center for Advanced Neuroimaging at University of California-Riverside. Anatomic images were acquired with an MP-RAGE sequence (echo time (TE)/repetition time (TR)/inversion time=3.02/2600/800 ms, flip angle=8°, voxel size=0.8×0.8×0.8 mm$^3$) for registration from subject space to common space. Multi-echo data were collected with a 12 echo 3D gradient recalled echo (GRE) sequence: TE$_1$/ΔTE/TR = 4/3/40 ms, FOV = 192 × 224 mm$^2$, matrix size of 192×224×96, slice thickness=1.7 mm, GRAPPA acceleration factor=2. Magnitude and phase images were saved for R$_2$* calculation and QSM processing, respectively. High-resolution diffusion-weighted MRI data were collected with a diffusion-weighted single-shot spin-echo, echo planar imaging sequence with the following parameters: TE / TR = 78 / 4500 ms, FOV = 194 × 168 mm$^2$, matrix size of 204 × 176, voxel size = 0.95 × 0.95 × 1 mm$^3$, multiband factor = 2, 64 slices with no gap, covering the brain from the middle of the cerebellum to the striatum. A monopolar diffusion encoding gradient was used to generate diffusion weighting. Diffusion-weighting gradients were applied in 30 directions with *b* values of 500 s/mm$^2$ and 2000 s/mm$^2$ with 32 *b*=0 images. Another set of 32 *b*=0 images with phase-encoding directions of opposite polarity were acquired to correct for susceptibility distortion [23].

The use of a monopolar diffusion encoding gradient was necessary to achieve spatial resolutions needed to resolve the LC. However, monopolar diffusion encoding gradients are particularly problematic in structures with elevated iron content, such as substantia nigra. Bipolar diffusion encoding gradients are insensitive to field inhomogeneities generated by iron [14, 24]. To examine the age-related effects without the influence of iron, an additional lower-resolution diffusion-weighted MRI dataset was collected with bipolar diffusion encoding gradients. Whole brain diffusion-weighted MRI data were collected with a diffusion-weighted single-shot spin-echo, echo planar imaging sequence with the following parameters: TE / TR = 102 / 3500 ms, FOV = 212 × 182 mm$^2$, matrix size of 128 × 110, voxel size = 1.7 × 1.7 × 1.7 mm$^3$, multiband factor = 4, 64 slices with no gap. Diffusion-weighting gradients were applied in 64 directions with *b* values of 1500 s/mm$^2$ and 3000 s/mm$^2$ with 3 *b*=0 images. For the diffusion-weighted MRI dataset with bipolar diffusion encoding gradients, two sets of diffusion-weighted images with phase-encoding directions of opposite polarity were acquired to correct for susceptibility distortion [23].

*2.2 Standard space transformation*

Imaging data were analyzed with FMRIB Software Library (FSL). A transformation was derived between individual subject space to Montreal Neurological Institute (MNI) 152 T$_1$-weighted space using FMRIB's Linear Image Registration Tool (FLIRT) and FMRIB's Nonlinear Image Registration Tool (FNIRT) in the FSL software package using the following steps [25, 26]. (1) The T$_1$-weighted image was skull stripped using the brain extraction tool (BET) in FSL, (2) brain extracted T$_1$-weighted images were aligned with the MNI brain extracted image using an affine transformation, and (3) a nonlinear transformation (FNIRT) was used to generate a transformation from individual T$_1$-weighted images to T$_1$-weighted MNI152 common space.

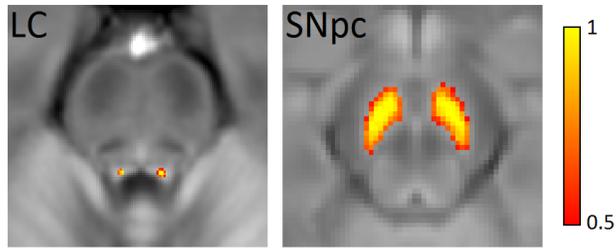

**Figure 1.** LC (left) and SNpc (right) masks used to define ROIs in the diffusion, $R_2^*$, and susceptibility analyses overlaid on mean NM-MRI images.

*2.3 Neuromelanin-sensitive atlases*

A SNpc neuromelanin atlas in MNI 152 space was created for use in this study from a separate cohort of 31 healthy participants (mean age: 26.1 years) using a process similar to those outlined earlier [27, 28]. The SNpc template was transferred from MNI 152 space to individual subject space using FLIRT and FNIRT in the FSL software package as follows [25, 26]. First, brain extracted $T_1$-weighted images were aligned with the MNI brain extracted image using an affine transformation. Second, a nonlinear transformation was used to generate a transformation from individual subject space to common space. Next, individual SNpc and masks were transformed to their respective $T_1$-weighted images using FLIRT and then transformed to common space using FNIRT. Finally, the first echo from the $T_2$-weighted GRE sequence was brain extracted using the brain extraction tool (BET) in FSL and a rigid body transformation was used to register the brain extracted $T_1$-weighted image with FLIRT. The resulting transformation matrix was used to transform the SNpc mask in $T_1$-space to $T_2$-weighted or diffusion space. The resultant SNpc masks were thresholded at a level of 0.6, corresponding to at least 60% of the subjects sharing the voxel, and then binarized. Each transformation was checked for errors, and no discernable difference was seen in the location of the ventricles in the $T_1$-weghted, $T_2$-weighted, and diffusion images.

An LC atlas was created from the same population used to create the SNpc neuromelanin atlas. LC was segmented using a previously used procedure [2], transformed to MNI standard space, and a population map was created using the steps outlined above. The LC atlas was thresholded at a level of 0.6, binarized, and then transformed to individual $T_2$-weighted or diffusion images as described above. LC and SNpc atlases used in this analysis are shown in Figure 1.

*2.4 Diffusion processing*

Diffusion-weighted data were analyzed with FSL [25, 29, 30] and MATLAB (The Mathworks, Natick, MA). Standard preprocessing steps were applied to correct susceptibility induced distortions in the diffusion MR data. Diffusion MR data were first corrected for eddy-current distortion, motion, and for susceptibility distortion using eddy in FSL [23, 31]. Next, skull stripping of the $T_1$-weighted image and susceptibility corrected $b=0$ image was performed using the brain extraction tool in the FSL software package [32]. Finally, measures derived from the diffusion MR data, including fractional anisotropy (FA) and mean diffusivity (MD) were calculated using the dtifit tool in FSL. Radial diffusivity (RD) and axial diffusivity (AD) was obtained from the resulting eigenvalue maps. Processing and registration steps are illustrated in Figure 2.

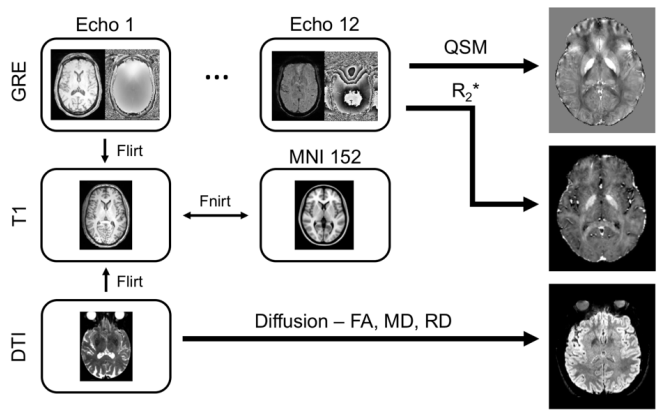

**Figure 2.** The registration and processing pipeline. QSM and $R_2$* images were created from the phase and magnitude images in the multi-echo GRE acquisition, respectively. FA, MD, and RD images were created from the high resolution diffusion-weighted acquisition. Registration between GRE and $T_1$ images was derived from the first echo of the GRE acquisition with a rigid body transformation, and registration between diffusion-weighted and $T_1$ images was derived using the $b=0$ image with a rigid body transformation with boundary based registration cost function.

*2.5 $R_2$* and QSM processing*

$R_2^*$ values were estimated using a custom script in MATLAB by fitting a monoexponential model to the GRE images.

$$S_i = S_0 \exp(-R_2^* TE) \quad [1]$$

where $S_0$ denotes a fitting constant and $S_i$ denotes the signal of a voxel at the $i$th echo time. The resulting $R_2$* map was aligned to the $T_1$-weighted image using a transform derived via the magnitude image from the first echo (FLIRT, degrees of freedom = 6). For each subject, mean $R_2$* was measured in the SNpc and in LC ROIs.

QSM images were constructed using the following procedure. Phase maps were constructed from the raw data, and phase offsets were removed. Next, phase maps were unwrapped [33]. The background phase was then removed with the spherical mean value method using a filter having a radius of eight pixels [34]. Finally, susceptibility maps were derived from the frequency map of brain tissue using an improved least-squares (iLSQR) method [35, 36] and Laplace filtering with a threshold of 0.04 as a truncation value. All QSM images were processed in MATLAB (The Math-Works, Inc., Natick, MA, USA) using in-house scripts. Mean susceptibility was measured in SNpc for each subject. Susceptibility was not measured in LC due to substantial susceptibility artifacts generated by the fourth ventricle.

*2.6 Statistical Analysis*

All statistical analyses were performed using IBM SPSS Statistics software version 24 (IBM Corporation, Somers, NY, USA) and results are reported as mean ± standard deviation. A *p* value of 0.05 was considered significant for all statistical tests performed in this work. Group $R_2^*$, susceptibility, and DTI indices comparisons between the young and aged cohort were made using separate two-tailed *t*-tests, and we expected group effects in each measure.

Spearman rank correlations of mean SNpc $R_2^*$ and mean SNpc susceptibility with age were performed separately in both cohorts. As histology suggests a rapid accrual of iron early in life and leveling off later in life [13], we expected significant correlations in mean SNpc $R_2^*$ and susceptibility in the young cohort but not in the aged cohort. The dependence of diffusivity on iron content in SNpc was ascertained with Spearman rank correlations between monopolar and bipolar diffusion indices and iron measures ($R_2^*$ or susceptibility), in each group. As the bipolar diffusion acquisition is insensitive to magnetic field inhomogeneities generated by iron, we predicted the diffusion indices from the monopolar acquisition would be sensitive to iron but not diffusion indices from the bipolar acquisition.

To assess the impact of LC microstructure on memory, we performed a correlation analysis between mean LC DTI indices and RAVLT delayed recall. Cognitive decline is linked to the loss of melanized neurons in LC [37] and we hypothesized that LC microstructural measures would correlate with RAVLT delayed recall score in the aged group.

## 3. Results

*3.1 LC microstructure*

Figure 3A shows comparison of mean FA in young (i) and aged (ii) groups. Age group differences in DTI indices and $R_2^*$ were assessed using between-group t-tests. Results revealed reductions in mean LC MD (aged: $3.30\times10^{-4}$ mm$^2$/s ±$4.27\times10^{-5}$ mm$^2$/s; young: $3.58\times10^{-4}$ mm$^2$/s ± $3.84\times10^{-5}$ mm$^2$/s; $t=2.52$; $p=0.007$), LC RD (aged: $2.73\times10^{-4}$ mm$^2$/s ±$3.93\times10^{-5}$ mm$^2$/s; young: $3.06\times10^{-4}$ mm$^2$/s ±$3.49\times10^{-5}$ mm$^2$/s; $t=2.64$; $p=0.005$), and LC AD (aged: $4.32\times10^{-4}$ mm$^2$/s ±$5.36\times10^{-5}$ mm$^2$/s; young: $4.57\times10^{-4}$ mm$^2$/s ±$4.41\times10^{-5}$ mm$^2$/s; $t=1.79$; $p=0.04$) in the aged cohort relative to the young. An increase in mean LC FA (aged: $0.29 \pm 0.04$; young: $0.27 \pm 0.02$; $t=-2.36$;

$p$=0.01) was observed in the aged cohort as compared to the young cohort. No difference was observed in mean $R_2^*$ in LC between the aged and young cohorts (aged: 18.2 s$^{-1}$ ± 1.9 s$^{-1}$; young: 17.4 s$^{-1}$ ± 1.8 s$^{-1}$; $t$=-0.88; $p$=0.07). Group comparisons for LC are summarized in Figure 3.

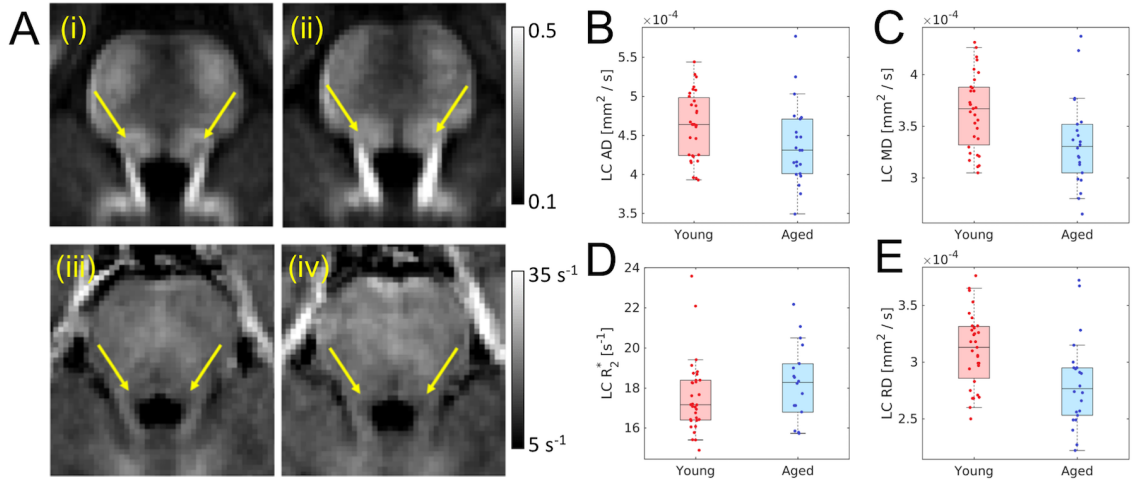

**Figure 3.** Part A shows comparison of mean LC FA in young (i) and aged (ii) cohorts as well as mean $R_2^*$ in young (iii) and aged (iv) cohorts. In A (i-iv) arrows show the approximate location of LC. Group comparisons are shown for AD (B), MD (C), and RD (D), and $R_2^*$ (E). A statistically significant increase in mean LC FA was observed in the aged cohort while reductions in mean LC RD and mean LC MD were seen in the aged cohort.

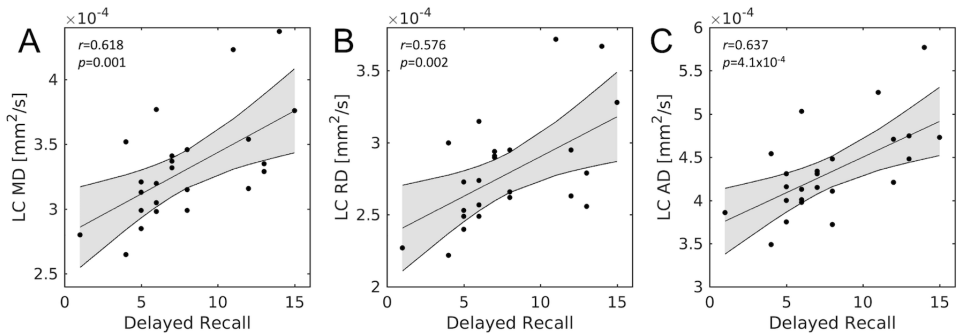

**Figure 4.** Correlations between mean LC diffusion measures and RAVLT delayed recall in the aged group. Statistically significant correlations were observed between mean LC MD (A), mean LC RD (B), and mean LC AD (C) with RAVLT delayed recall.

*3.2 LC microstructure and memory*

The effect of LC microstructure on memory was assessed by correlating LC diffusion indices (MD, RD, and AD) with RVALT delayed recall components. In the aged group mean LC diffusion metrics positively correlate with memory scores (RAVLT delayed recall), indicating that as memory performance increases, LC MD ($r=0.618$, $p=0.001$), mean LC RD ($r=0.576$, $p=0.002$), and mean LC AD ($r=0.637$, $p=4.1\times10^{-4}$) also increase. These correlations are shown in Figure 4.

Interestingly, no correlation between any LC diffusion measures and RAVLT delayed recall score was observed in the young group (mean LC MD: $r=0.074$, $p=0.487$; mean LC RD: $r=0.006$, $p=0.346$; mean LC AD: $r=0.119$, $p=0.262$).

*3.3 SNpc microstructure and composition*

SNpc microstructure was assessed using iron sensitive (monopolar; high resolution) and iron insensitive (bipolar; lower resolution with 1.7 mm isotropic voxel) diffusion-weighted acquisitions. Figure 5 shows mean SNpc FA from the high resolution (iron sensitive) acquisition in young (i) and aged (ii) cohorts, respectively. In the iron sensitive acquisition, reductions in mean SNpc MD ($t=3.22$; $p=0.002$), mean SNpc RD ($t=2.98$; $p=0.004$), and mean SNpc AD ($t=3.09$; $p=0.002$) were observed in the aged cohort as compared to the young. A trend toward increased FA was observed in the aged cohort ($t=-1.43$; $p=0.08$). In the iron insensitive diffusion acquisition, in the aged cohort, reductions in MD ($t=2.57$; $p=0.009$) and AD ($t=3.07$; $p=0.002$) were observed and trends toward decreased RD ($t=2.17$; $p=0.10$) and decreased FA ($t=-1.37$; $p=0.06$) were seen. Group comparisons for SNpc diffusion markers are detailed in Table 2.

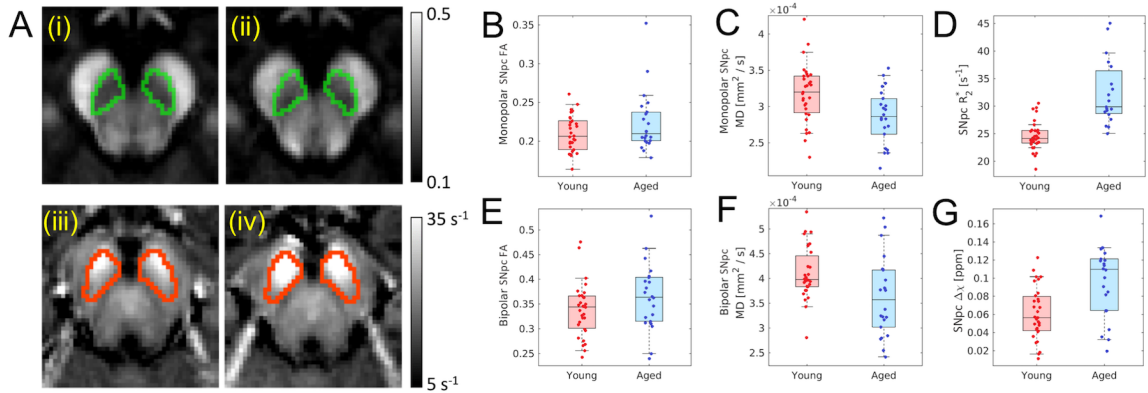

**Figure 5.** Part A shows comparison of mean SNpc FA in young (i) and aged (ii) cohorts as well as mean $R_2^*$ in young (iii) and aged (iv) cohorts. SNpc is outlined in green in mean FA maps (i and ii) and red in mean $R_2^*$ maps (iii and iv). Group comparisons are shown for monopolar FA (B) and bipolar FA (E), monopolar MD (C) and bipolar MD (F), $R_2^*$ (D), and susceptibility (G). Statistically significant increases in mean SNpc $R_2^*$ and susceptibility were observed in the aged cohort while a reduction in mean SNpc MD were seen in the aged cohort for monopolar and bipolar DTI acquisitions.

Changes in SNpc composition were assessed using iron sensitive contrasts, and Figure 5 (iii) and (iv) show mean SNpc $R_2^*$ in young and aged cohorts, respectively. An increase in $R_2^*$ (aged: 32.4 s$^{-1}$ ± 5.9s$^{-1}$; young: 24.6 s$^{-1}$ ±2.6 s$^{-1}$; $t=-5.87$; $p<10^{-5}$) as well as susceptibility (aged: 0.096 ppm ± 0.04 ppm; young: 0.06 ppm ± 0.03 ppm; $t$=-4.09; $p$=1.0×10$^{-4}$) was observed in SNpc of the aged cohort.

No correlations between age and diffusion indices in SNpc were observed in the aging cohort (FA: $p$=0.656; MD: $p$=0.925; RD: $p$=0.912) or in the young cohort (FA: $p$=0.481; MD: $p$=0.834; RD: $p$=0.676). A significant correlation was observed between iron in SNpc and age in the young cohort ($R_2^*$: $p$=0.008, $r$=0.462; susceptibility: $p$=0.01, $r$=0.445) but no correlation was seen between SNpc iron and age ($R_2^*$: $p$=0.99, $r$=0.004; susceptibility: $p$=0.877, $r$=-0.111) in the aged cohort.

**Table 2.** Summary of SNpc diffusion markers from the monopolar (iron sensitive) and bipolar diffusion acquisitions (iron insensitive). Data are presented as mean ± standard deviation MD – mean diffusivity; AD – axial diffusivity; RD – radial diffusivity; FA – fractional anisotropy.

| SNpc Marker | Monopolar Diffusion Encoding Gradient | | | Bipolar Diffusion Encoding Gradient | | |
|---|---|---|---|---|---|---|
| | Young | Aged | $p$ Value | Young | Aged | $p$ Value |
| MD [mm$^2$/s] | 3.2×10$^{-4}$±4.1×10$^{-5}$ | 2.9×10$^{-4}$±3.8×10$^{-5}$ | 0.001 | 4.1×10$^{-4}$ ± 5.3×10$^{-5}$ | 3.6×10$^{-4}$±8.0×10$^{-5}$ | 0.009 |
| AD [mm$^2$/s] | 3.4×10$^{-4}$±4.5×10$^{-5}$ | 3.7×10$^{-4}$±3.8×10$^{-5}$ | 0.002 | 5.6×10$^{-4}$ ±6.4×10$^{-5}$ | 4.8×10$^{-4}$ ±9.8×10$^{-5}$ | 0.002 |
| RD [mm$^2$/s] | 2.8×10$^{-4}$±3.7×10$^{-5}$ | 2.5×10$^{-4}$±3.6×10$^{-5}$ | 0.004 | 3.4×10$^{-4}$ ±5.2×10$^{-5}$ | 3.0×10$^{-4}$ ±7.6×10$^{-5}$ | 0.10 |
| FA | 0.29 ± 0.04 | 0.27 ± 0.02 | 0.06 | 0.34 ± 0.05 | 0.36 ± 0.07 | 0.06 |

*3.4 Effect of iron of SNpc diffusion measures*

Statistically significant correlations were found in the aged group between iron measures and diffusion measures in SNpc from both acquisitions. Specifically, negative correlations were seen in the aged group between mean SNpc $R_2^*$ all SNpc diffusion measures from the monopolar acquisition (MD: $r$=-0.638, $p$=0.001; AD: $r$=-0.626, $p$=0.001; RD: $r$=-0.661, $p$=0.001) and those from the bipolar acquisition (MD: $r$=-0.743, $p$<10$^{-4}$; AD: $r$=-0.749; $p$<10$^{-4}$; RD: $r$=-0.660, $p$=0.001). These correlations are shown in Figure 6. Correlations of similar strength were observed between mean SNpc Δχ and all SNpc diffusion measures from the monopolar acquisition (MD: $r$=-0.592, $p$=0.002; AD: $r$=-0.586, p=0.002; RD: $r$=-0.611, p=0.002) and those from the bipolar acquisition (MD: $r$=-0.782, $p$<10$^{-4}$; AD: $r$=-0.724; $p$<10$^{-4}$; RD: $r$=-0.772, $p$<10$^{-4}$).

In the young group, strong negative trends were observed between mean SNpc $R_2^*$ and all SNpc diffusion measures from the monopolar acquisition (MD: $r$=-0.278, $p$=0.055; AD: $r$=-0.245, $p$=0.082; RD: $r$=-0.259, p=0.070) but no trend was seen between mean SNpc $R_2^*$ and any SNpc diffusion measure from the bipolar acquisition (MD: $r$=-0.025, $p$=0.444; AD: $r$=-0.071, $p$=0.344; RD: $r$=0.127, $p$=0.238). However, for SNpc susceptibility, mean SNpc Δχ showed strong negative trends with all SNpc diffusion measures from the monopolar acquisition (MD: $r$=-0.231, $p$=0.085;

AD: *r*=-0.207, *p*=0.118; RD: *r*=-0.279, *p*=0.065) as well as with SNpc diffusion measures from the bipolar acquisition (MD: *r*=-0.432, *p*=0.011; AD: *r*=-0.325, *p*=0.026; RD: *r*=-0.419, *p*=0.070).

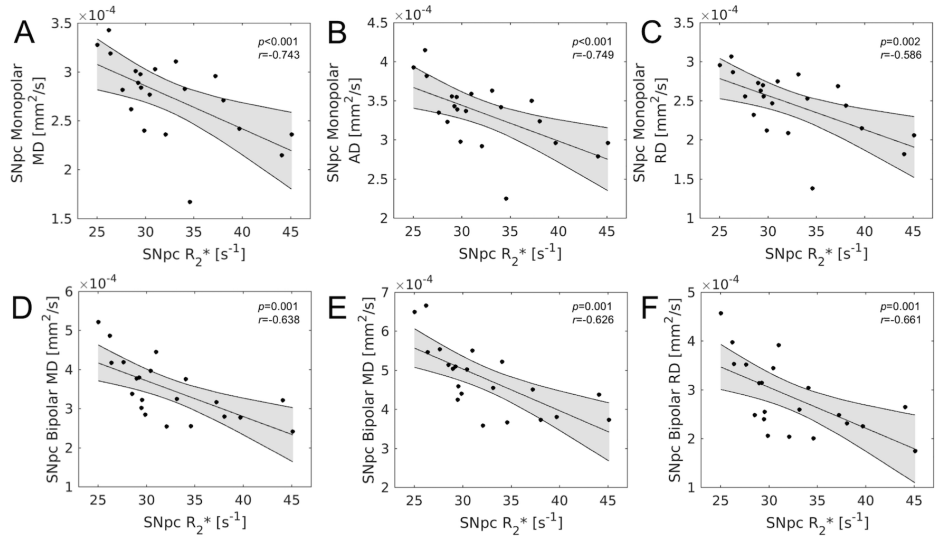

**Figure 6.** Correlations between $R_2^*$ and diffusion measures in SNpc in the aged group. Statistically significant correlations were observed between all bipolar diffusion measures (A – SNpc MD, B – SNpc AD, C- SNpc RD) and $R_2^*$ as well as between all monopolar diffusion measures (D – SNpc MD, E – SNpc AD, F- SNpc RD) and $R_2^*$.

## 4. Discussion

This study examined the microstructural and compositional differences occurring in catecholamine nuclei during normal aging. In the LC, aging led to expected age-related reductions in diffusivity (MD, RD) and an unexpected increase in anisotropy (FA), but there was no effect of aging on $R_2^*$. Aging in SNpc was accompanied by reductions in diffusivity (but not anisotropy) with additional increases in measures of iron deposition.

The stature of LC is too small for assessment of microstructural changes using traditional diffusion weighted acquisitions, so a high-resolution diffusion weighted acquisition was employed to minimize partial volume effects. We observed increased FA, decreased MD, and decreased RD in LC of the aging cohort as compared to controls. Although diffusion measures can be altered by several microstructural properties, we conjecture that the observed diffusion changes reflect restricted diffusion occurring from a decrease in axon diameter [38,39]. Prior postmortem studies examining LC have found a reduction in LC cell size of aged humans [40]. Interestingly, the trend toward increased LC $R_2^*$ in the aged cohort may reflect reduced cell size as tissue compacting or shrinkage increases $R_2^*$. The age-related increase in LC $R_2^*$ cannot be attributed to iron deposition as LC neuromelanin granules chelate copper [41] and neuromelanin is only slightly paramagnetic [42].

LC is the primary source of norepinephrine in the brain, and its projections innervate many areas of the brain including regions deeply involved in learning and memory, such as the neocortex and hippocampus [43]. Norepinephrine protects neurons from oxidative stress [44] and reduces inflammation [45]. This protective effect is the basis of the LC-reserve hypothesis by which activation of LC and the release of norepinephrine may offer neuroprotection to neurons innervated by LC [46]. Histological studies found that higher density of neurons in LC correlates with a reduction in cognitive decline [37] and an earlier study found a link between LC contrast and the retention of cognitive reserve [47]. Interestingly, we found that lower performance on RAVLT delayed recall scores were correlated with reduced MD and RD (i.e. reduced LC axon size), suggesting that LC microstructure integrity may be important in the retention of cognitive abilities during aging.

Histology has found an age dependence for iron load in substantia nigra [13], which has been confirmed in several imaging studies [17-20]. Increases in substantia nigra susceptibility, $R_2^*$, or field-dependent relaxation rates have been observed in cohorts of aged subjects as compared to young subjects [18, 20] and the results presented here accord with these findings. Other work examining iron deposition in a wider age range has been inconclusive with one study finding that iron increases in substantia nigra until early adulthood and then plateaus [17] while another showed iron load increases throughout life [19]. Interestingly, results presented here agree with both studies. Specifically, significant positive correlations were seen between SNpc iron measures and age in the young cohort but no significant correlations in iron measures were observed in the aged cohort, suggesting high accrual rates early in life while group effects suggest iron accrual continues throughout life.

Determination of age-related iron deposition in substantia nigra or its subcomponents (SNpc or substantia nigra pars reticulata) is essential for the establishment of biomarkers to distinguish normal $R_2^*$ or QSM values from those due to a pathological condition. For example, in Parkinson's disease, much of the neuronal loss in SNpc occurs prior to symptom onset and increased iron content in substantia nigra is associated with this neuronal loss [48, 49]. However, imaging studies have not reached a consensus regarding nigral iron deposition in Parkinson's disease [50-53], and this inconsistency may be due to placement of substantia ROIs outside SNpc [54].

Iron is particularly problematic for diffusion-weighted MRI sequences with monopolar diffusion encoding gradients. Deposits of iron create local magnetic field gradients, which produce cross terms affecting monopolar diffusion encoding gradients and reduce the apparent diffusion coefficient [14, 24]. Our results from the monopolar diffusion-weighted acquisition accord with this physical model. We found negative correlations between both SNpc iron measures (mean $R_2^*$ and mean susceptibility) and diffusivity (MD, AD, and RD) in data acquired with a monopolar diffusion encoding gradient in the aged group. However, similar correlations were observed in the diffusion-weighted data acquired with a bipolar diffusion encoding gradient. This latter finding is particularly puzzling since physical models suggest bipolar diffusion encoding gradients are insensitive to magnetic field inhomogeneities generated by iron deposits [15].

Negative correlations between iron measures and MD as well as positive correlations between FA and iron measures have been observed in the striatum [12, 55]. These results suggest a complex relationship between iron deposition and diffusivity measures that warrants further study. Specifically, correlations between SNpc susceptibility and diffusion measures from the bipolar acquisition in both groups suggest iron deposition may alter the underlying tissue microstructure or that the correlation between microstructure and iron are incidental and age-related microstructural changes are dominant. Characterization of the relationship between iron and diffusivity is of particular interest in deep gray matter nuclei undergoing age-related or pathologic iron deposition. In particular, this is of interest in the study of PD where iron is deposited in SNpc [56]. The development of diagnostic and progression markers for PD has been hindered by inconsistent results from different studies [57] and discrepancies in the literature may be partially explained by the sensitivity of diffusion measures to iron content.

Given the negative correlations for diffusion measures and the increase in SNpc iron measures in the aged cohort, it is unsurprising that reduced diffusivity was observed in the aged cohort. Earlier work examining age-related microstructural differences in substantia nigra found a trend towards reduced MD ($p$=0.12), a statistically significant reduction in FA, and a statistically significant increase in RD in their aged cohort [21]. Differences in results between our study and the previous report [21] cannot be attributed to a discrepancy in ROI placement. ROIs were drawn in hypointense regions inferior to red nucleus and were positioned in a spatial location similar to the rostral portion of the neuromelanin-sensitive SN (SNpc) ROI used in this analysis [22]. Given this, the discrepancy in RD and FA results may be attributed to partial volume effects as their acquisition had a slightly larger voxel size than the diffusion acquisition used in this work.

There are some caveats in the present study. First, although a neuromelanin-sensitive atlas created by our group was applied to these data, neuromelanin-sensitive data were not acquired in the same cohort. Histological data suggests that neuromelanin may be present in LC, but not substantia nigra, at birth [58] and neuromelanin accrues in both structures with age [59]. Neuromelanin content in both structures peaks around age 60 and decreases later in life [9, 59, 60] with depigmentation of substantia nigra occurring at a rate of 4.7% loss per decade of life [10]. Inclusion of neuromelanin-sensitive data would give a holistic assessment of age-related effects. Second, the age distribution of the subjects in this study was bimodal, owing to its cross-sectional design, with the mean ages of 20.9 years and 73.5 years for the two populations studied. No subjects between the ages of 26 and 60 participated in this study. Third, it is possible that noise may corrupt measurements in SNpc from the high-resolution (monopolar) DTI acquisition. Low SNR from iron deposition or reduced voxel size will negatively bias radial diffusivity but positively bias axial diffusivity, resulting in an increase in FA [61]. We speculate that the effect of noise on the monopolar acquisition is minimal since similar effect sizes are seen in both DTI acquisitions since SNpc SNR in the bipolar acquisition is relatively high (mean SNpc SNR = 82.3). However, additional work is needed to assess the contribution of noise on diffusion indices in iron containing structures. Finally, although participants were screened for self-reported neurological conditions prior to enrollment, it is impossible to exclude the possibility that some of the healthy controls in the aged cohort have an undiagnosed neurological condition. As mentioned above, LC and SNpc are profoundly effected in the prodromal stages of Alzheimer's disease (LC only) or PD (LC and SNpc) [7, 8].

## 5. Conclusion

In this work, a high-resolution diffusion-weighted MRI protocol and a multi-echo gradient echo protocol were employed to examine age-related microstructural and compositional differences in LC and SNpc. Older age was most associated with axonal thinning in LC whereas age-related differences in SNpc are likely due to increased iron content. Further, these differences in iron content were found to strongly correlate with diffusion measurements in SNpc in the aged cohort while strong trends were observed between susceptibility and diffusion in the young cohort.

## Acknowledgements

Xiaoping Hu receives support from the Michael J. Fox Foundation (MJF 10854). This work was also supported by R00 AG047334 (Bennett) and R21 AG054804 (Bennett) from the National Institutes of Health/National Institute on Aging. The authors would like to thank Mrs. Chelsea Evelyn for help with data acquisition.

## References

1. Sasaki M, Shibata E, Tohyama K, et al. Neuromelanin magnetic resonance imaging of locus ceruleus and substantia nigra in Parkinson's disease. Neuroreport 2006;17(11):1215-1218.

2. Chen X, Huddleston DE, Langley J, et al. Simultaneous imaging of locus coeruleus and substantia nigra with a quantitative neuromelanin MRI approach. Magn Reson Imaging 2014;32(10):1301-1306.

3. Schwarz ST, Rittman T, Gontu V, Morgan PS, Bajaj N, Auer DP. T1-Weighted MRI shows stage-dependent substantia nigra signal loss in Parkinson's disease. Mov Disord 2011;26(9):1633–1638.

4. Carter ME, Yizhar O, Chikahisa S, et al. Tuning arousal with optogenetic modulation of locus coeruleus neurons. Nat Neurosci 2010;13(12):1526-1533.

5. Aston-Jones G, Cohen JD. Adaptive gain and the role of the locus coeruleus–norepinephrine system in optimal performance. J Comp Neurol 2005;493(1):99-110.

6. Jahanshahi M, Jones CR, Dirnberger G, Frith CD. The substantia nigra pars compacta and temporal processing. J Neurosci 2006;26(47):12266-12273.

7. Braak H, Thal DR, Ghebremedhin E, Del Tredici K. Stages of the pathologic process in Alzheimer disease: age categories from 1 to 100 years. Journal of neuropathology and experimental neurology 2011;70(11):960-969.

8. Braak H, Tredici KD, Rüb U, de Vos RAI, Jansen Steur ENH, Braak E. Staging of brain pathology related to sporadic Parkinson's disease. Neurobiol Aging 2003;24(2):197-211.

9. Manaye KF, McIntire DD, Mann DM, German DC. Locus coeruleus cell loss in the aging human brain: a non-random process. J Comp Neurol 1995;358(1):79-87.

10. Fearnley JM, Lees AJ. Ageing and Parkinson's disease: substantia nigra regional selectivity. Brain 1991;114 ( Pt 5):2283-2301.

11. Beaulieu C. The basis of anisotropic water diffusion in the nervous system - a technical review. NMR Biomed 2002;15(7-8):435-455.

12. Pfefferbaum A, Adalsteinsson E, Rohlfing T, Sullivan EV. Diffusion tensor imaging of deep gray matter brain structures: effects of age and iron concentration. Neurobiol Aging 2010;31(3):482-493.

13. Hallgren B, Sourander P. The effect of age on the non-haemin iron in the human brain. J Neurochem 1958;3(1):41-51.

14. Zhong J, Kennan RP, Gore JC. Effects of susceptibility variations on NMR measurements of diffusion. J Magn Reson 1991;95:267-280.

15. Fujiwara S, Uhrig L, Amadon A, Jarraya B, Le Bihan D. Quantification of iron in the non-human primate brain with diffusion-weighted magnetic resonance imaging. Neuroimage 2014;102(2):789-797.

16. Langkammer C, Schweser F, Krebs N, et al. Quantitative susceptibility mapping (QSM) as a means to measure brain iron? A post mortem validation study. Neuroimage 2012;62(3):1593-1599.

17. Aquino D, Bizzi A, Grisoli M, et al. Age-related iron deposition in the basal ganglia: quantitative analysis in healthy subjects. Radiology 2009;252(1):165-172.

18. Bilgic B, Pfefferbaum A, Rohlfing T, Sullivan EV, Adalsteinsson E. MRI estimates of brain iron concentration in normal aging using quantitative susceptibility mapping. Neuroimage 2012;59(3):2625-2635.

19. Haacke EM, Miao Y, Liu M, et al. Correlation of putative iron content as represented by changes in R2* and phase with age in deep gray matter of healthy adults. J Magn Reson Imaging 2010;32(3):561-576.

20. Pfefferbaum A, Adalsteinsson E, Rohlfing T, Sullivan EV. MRI estimates of brain iron concentration in normal aging: comparison of field-dependent (FDRI) and phase (SWI) methods. Neuroimage 2009;47(2):493-500.

21. Vaillancourt DE, Spraker MB, Prodoehl J, Zhou XJ, Little DM. Effects of aging on the ventral and dorsal substantia nigra using diffusion tensor imaging. Neurobiol Aging 2012;33(1):35-42.

22. Langley J, Huddleston DE, Chen X, Sedlacik J, Zachariah N, Hu X. A multicontrast approach for comprehensive imaging of substantia nigra. Neuroimage 2015;112:7-13.

23. Andersson JL, Skare S, Ashburner J. How to correct susceptibility distortions in spin-echo echo-planar images: application to diffusion tensor imaging. NeuroImage 2003;20(2):870-888.

24. Novikov DS, Reisert M, Kiselev VG. Effects of mesoscopic susceptibility and transverse relaxation on diffusion NMR. J Magn Reson 2018;293:134-144.

25. Smith SM, Jenkinson M, Woolrich MW, et al. Advances in functional and structural MR image analysis and implementation as FSL. NeuroImage 2004;23 Suppl 1:S208-219.

26. Woolrich MW, Jbabdi S, Patenaude B, et al. Bayesian analysis of neuroimaging data in FSL. NeuroImage 2009;45(1 Suppl):S173-186.

27. Langley J, Huddleston DE, Merritt M, et al. Diffusion tensor imaging of the substantia nigra in Parkinson's disease revisited. Hum Brain Mapp 2016;37(7):2547-2556.

28. Langley J, Huddleston DE, Sedlacik J, Boelmans K, Hu XP. Parkinson's disease-related increase of T2*-weighted hypointensity in substantia nigra pars compacta. Mov Disord 2017;32(3):441-449.

29. Jenkinson M, Bannister P, Brady M, Smith S. Improved optimization for the robust and accurate linear registration and motion correction of brain images. NeuroImage 2002;17(2):825-841.

30. Jenkinson M, Smith S. A global optimisation method for robust affine registration of brain images. Med Image Anal 2001;5(2):143-156.

31. Andersson JLR, Sotiropoulos SN. An integrated approach to correction for off-resonance effects and subject movement in diffusion MR imaging. Neuroimage 2016;125:1063-1078.

32. Smith SM. Fast robust automated brain extraction. Hum Brain Mapp 2002;17(3):143-155.

33. Langley J, Zhao Q. Unwrapping magnetic resonance phase maps with Chebyshev polynomials. Magn Reson Imaging 2009;27(9):1293-1301.

34. Schweser F, Deistung A, Lehr BW, Reichenbach JR. Quantitative imaging of intrinsic magnetic tissue properties using MRI signal phase: an approach to in vivo brain iron metabolism? NeuroImage 2011;54(4):2789-2807.

35. Li W, Wu B, Liu C. Quantitative susceptibility mapping of human brain reflects spatial variation in tissue composition. NeuroImage 2011;55(4):1645-1656.

36. Li W, Wang N, Yu F, et al. A method for estimating and removing streaking artifacts in quantitative susceptibility mapping. Neuroimage 2015;108:111-122.

37. Wilson RS, Nag S, Boyle PA, et al. Neural reserve, neuronal density in the locus ceruleus, and cognitive decline. Neurology 2013;80(13):1202-1208.

38. Barazany D, Basser PJ, Assaf Y. In vivo measurement of axon diameter distribution in the corpus callosum of rat brain. Brain 2009;132(Pt 5):1210-1220.

39. Wheeler-Kingshott CA, Cercignani M. About "axial" and "radial" diffusivities. Magn Reson Med 2009;61(5):1255-1260.

40. German DC, Manaye KF, White CL, 3rd, et al. Disease-specific patterns of locus coeruleus cell loss. Ann Neurol 1992;32(5):667-676.

41. Sulzer D, Cassidy C, Horga G, et al. Neuromelanin detection by magnetic resonance imaging (MRI) and its promise as a biomarker for Parkinson's disease. NPJ Parkinsons Dis 2018;4:11.

42. Ju KY, Lee JW, Im GH, et al. Bio-inspired, melanin-like nanoparticles as a highly efficient contrast agent for T1-weighted magnetic resonance imaging. Biomacromolecules 2013;14(10):3491-3497.

43. Aston-Jones G, Cohen JD. An integrative theory of locus coeruleus-norepinephrine function: adaptive gain and optimal performance. Annu Rev Neurosci 2005;28:403-450.

44. Troadec JD, Marien M, Darios F, et al. Noradrenaline provides long-term protection to dopaminergic neurons by reducing oxidative stress. J Neurochem 2001;79(1):200-210.

45. Feinstein DL, Heneka MT, Gavrilyuk V, Dello Russo C, Weinberg G, Galea E. Noradrenergic regulation of inflammatory gene expression in brain. Neurochem Int 2002;41(5):357-365.

46. Robertson IH. A noradrenergic theory of cognitive reserve: implications for Alzheimer's disease. Neurobiol Aging 2013;34(1):298-308.

47. Clewett DV, Lee TH, Greening S, Ponzio A, Margalit E, Mather M. Neuromelanin marks the spot: identifying a locus coeruleus biomarker of cognitive reserve in healthy aging. Neurobiol Aging 2016;37:117-126.

48. Dexter DT, Carayon A, Javoy-Agid F, et al. Alterations in the levels of iron, ferritin and other trace metals in Parkinson's disease and other neurodegenerative diseases affecting the basal ganglia. Brain 1991;114 ( Pt 4):1953-1975.

49. Wypijewska A, Galazka-Friedman J, Bauminger ER, et al. Iron and reactive oxygen species activity in parkinsonian substantia nigra. Parkinsonism & related disorders 2010;16(5):329-333.

50. Heim B, Krismer F, De Marzi R, Seppi K. Magnetic resonance imaging for the diagnosis of Parkinson's disease. J Neural Transm (Vienna) 2017;124(8):915-964.

51. Lehericy S, Vaillancourt DE, Seppi K, et al. The role of high-field magnetic resonance imaging in parkinsonian disorders: Pushing the boundaries forward. Mov Disord 2017;32(4):510-525.

52. Du G, Lewis MM, Sica C, et al. Distinct progression pattern of susceptibility MRI in the substantia nigra of Parkinson's patients. Mov Disord 2018.

53. Huddleston DE, Langley J, Dusek P, et al. Imaging parkinsonian pathology in substantia nigra with MRI. Curr Radiol Rep 2018;6:15.

54. Langley J, He N, Huddleston DE, et al. Reproducible detection of nigral iron deposition in 2 Parkinson's disease cohorts. Mov Disord 2019;34(3):416-419.

55. Syka M, Keller J, Klempir J, et al. Correlation between relaxometry and diffusion tensor imaging in the globus pallidus of Huntington's disease patients. PLoS One 2015;10(3):e0118907.

56. Dexter DT, Wells FR, Agid F, et al. Increased nigral iron content in postmortem parkinsonian brain. Lancet 1987;2(8569):1219-1220.

57. Schwarz ST, Abaei M, Gontu V, Morgan PS, Bajaj N, Auer DP. Diffusion tensor imaging of nigral degeneration in Parkinson's disease: A region-of-interest and voxel-based study at 3 T and systematic review with meta-analysis. Neuroimage Clin 2013;3:481-488.

58. Fenichel GM, Bazelon M. Studies on neuromelanin. II. Melanin in the brainstems of infants and children. Neurology 1968;18(8):817-820.

59. Zecca L, Stroppolo A, Gatti A, et al. The role of iron and copper molecules in the neuronal vulnerability of locus coeruleus and substantia nigra during aging. Proc Natl Acad Sci U S A 2004;101(26):9843-9848.

60. Ma SY, Roytt M, Collan Y, Rinne JO. Unbiased morphometrical measurements show loss of pigmented nigral neurones with ageing. Neuropathol Appl Neurobiol 1999;25(5):394-399.

61. Anderson AW. Theoretical analysis of the effects of noise on diffusion tensor imaging. Magn Reson Med 2001;46(6):1174-1188.